\documentclass{osa-article}

\journal{osajournal}
\articletype{Research Article}
\begin{document}

\title{Optimal illumination scheme for isotropic quantitative differential phase contrast microscopy}

\author{Yao Fan,\authormark{1,2,3,5} Jiasong Sun,\authormark{1,2,3,5} Qian Chen,\authormark{1,2,*} Xiangpeng Pan,\authormark{1,2,3}, Lei Tian,\authormark{4} and Chao Zuo,\authormark{1,2,3,*}}

\address{\authormark{1}School of Electronic and Optical Engineering, Nanjing University of Science and Technology, No. 200 Xiaolingwei Street, Nanjing, Jiangsu Province 210094, China \\
\authormark{2}Jiangsu Key Laboratory of Spectral Imaging $\&$ Intelligent Sense, Nanjing University of Science and Technology, Nanjing, Jiangsu Province 210094, China\\
\authormark{3}Smart Computational Imaging (SCI) Laboratory, Nanjing University of Science and Technology, Nanjing, Jiangsu Province 210094, China\\
\authormark{4}Department of Electrical and Computer Engineering, Boston University, Boston, MA 02215, USA \\
\authormark{5}These authors contributed equally to this work.\\
}

\email{\authormark{*}chenqian@njust.edu.cn\\
       \authormark{*} zuochao@njust.edu.cn\\} %% email address is required



\begin{abstract}
   Differential phase contrast microscopy (DPC) provides high-resolution quantitative phase distribution of thin transparent samples under multi-axis asymmetric illuminations. Typically, illumination in DPC microscopic systems is designed with 2-axis half-circle amplitude patterns, which, however, result in a non-isotropic phase contrast transfer function (PTF). Efforts have been made to achieve isotropic DPC by replacing the conventional half-circle illumination aperture with radially asymmetric patterns with 3-axis illumination or gradient amplitude patterns with 2-axis illumination. Nevertheless, these illumination apertures were empirically designed based on empirical criteria related to the shape of the PTF, leaving the underlying theoretical mechanisms unexplored. Furthermore, the frequency responses of the PTFs under these engineered illuminations have not been fully optimized, leading to suboptimal phase contrast and signal-to-noise ratio (SNR) for phase reconstruction. In this Letter, we provide a rigorous theoretical analysis about the necessary and sufficient conditions for DPC to achieve perfectly isotropic PTF. In addition, we derive the optimal illumination scheme to maximize the frequency response for both low and high frequencies (from 0 to $2NA_{obj}$), and meanwhile achieve perfectly isotropic PTF with only 2-axis intensity measurements. We present the derivation, implementation, simulation and experimental results demonstrating the superiority of our method over state-of-the-arts in both phase reconstruction accuracy and noise-robustness.
\end{abstract}

%%%%%%%%%%%%%%%%%%%%%%%%%%  body  %%%%%%%%%%%%%%%%%%%%%%%%%%
 \section{Introduction}
 \noindent
Quantitative phase imaging (QPI), which provides phase information about the refractive index distribution of transparent specimens, has drawn much attention in both optical and biomedical research \cite{popescu2011quantitative, barty1998quantitative,cuche1999digital}. The major advantage of QPI over conventional intensity imaging or fluorescence microscopy is that it requires no exogenous contrast agents (e.g., dyes or fluorescent protein) to enhance the contrast of the microscopic image, which enables label-free and stain-free optical imaging of live biological specimens \emph{in vitro} \cite{kim2014profiling,popescu2008quantitative}. The most common QPI methods are based on interferometry and holography with coherent illumination and a reference beam, making them expensive and sensitive to misalignment, vibrations, and speckle noise \cite{mann2005high,marquet2005digital,kemper2008digital}. To overcome these limitations, non-interferometric QPI approaches using partially coherent illumination have been developing, such as transport-of-intensity equation (TIE) \cite{teague1983deterministic,barty1998quantitative,kou2010transport,petruccelli2013transport,zuo2013high,zuo2013noninterferometric,zuo2013transport}, differential phase contrast (DPC) \cite{pfeiffer2006phase,hamilton1984differential,kachar1985asymmetric,mehta2009quantitative,tian20143d,tian2015quantitative}, and Fourier ptychographic microscopy (FPM) \cite{zheng2013wide,ou2013quantitative,tian2014multiplexed,zuo2016adaptive,sun2017resolution}. TIE is a well-established non-interferometric phase retrieval approach, which enables QPI of transparent sample simply by measuring the intensities at multiple axially displaced planes \cite{zuo2013transport,barty1998quantitative,petruccelli2013transport,zuo2013high,teague1983deterministic,zuo2013noninterferometric}. The advantages of TIE approach are that it is fully compatible with widely available bright-field microscopy hardware, and able to offer an imaging resolution up to the incoherent resolution limit ($2\times$ better than the coherent diffraction limit) under matched annular illumination \cite{zuo2017high,li2018optimal}. Without moving the position of a sample, DPC and FPM approaches retrieve the complex field of the sample by using asymmetric illuminations. In FPM, a set of low-resolution (LR) intensity images corresponding to different illumination angles, with the resolution determined by the numerical aperture (NA) of the objective lens, is acquired \cite{zheng2013wide,ou2013quantitative,zuo2016adaptive,sun2017resolution}. These LR intensity images are iteratively combined together in the Fourier domain, resulting in a wide-field, high-resolution complex image with the synthesized resolution determined by the sum of the objective lens and illumination NAs   \cite{zheng2013wide,sun2017resolution,ou2013quantitative,zuo2016adaptive,tian2014multiplexed,sun2018high,sun2016efficient}. DPC, however, achieves phase recovery by using only four images with asymmetric illuminations in opposite directions \cite{tian20143d,mehta2009quantitative}. It converts invisible sample phases into measurable intensity by shifting the sample's spectrum in Fourier space to theoretically achieve a resolution of twice the coherent diffraction limit \cite{tian20143d,mehta2009quantitative}. Assuming a linearized model for a weakly scattering sample, the DPC phase retrieval problem becomes a single-step deconvolution process using the phase contrast transfer function (PTF). By implementing DPC with a programmable LED array  \cite{tian20143d,tian2015quantitative,lee2015color} or a programmable LCD panel \cite{zuo2016programmable}, we are able to realize dynamic QPI along arbitrary axes of asymmetry, without any mechanical moving parts.

In order to recover the quantitative phase information of a weakly scattering sample, at least two complementary source patterns are required in DPC. However, the resultant PTF is anti-symmetric and zero at all spatial frequencies along the axis of asymmetry, which may lead to significant phase reconstruction artifacts if not properly handled. So in general, the illumination of DPC is designed with 2-axis half-circle amplitude patterns, i.e., 4 patterns (top, bottom, left, right half-circles) are used to avoid missing frequencies. However, artifacts of phase reconstruction still cannot be completely avoided, since DPC's PTF is not circularly symmetric with only 2-axis measurements. Efforts have been made toward developing high speed, or even single-shot QPI mechanisms, such as color-coded DPC based on wavelength multiplexing \cite{lee2015color,lee2017single,phillips2017single}. Other efforts have been made to achieve isotropic DPC by either using more time-consuming intensity measurements along different illumination angles (12-axis) \cite{tian2015quantitative} or replacing the conventional half-circle illumination aperture with 3-axis radially asymmetric patterns  \cite{lin2018quantitative} or 2-axis gradient amplitude patterns \cite{chen2018isotropic}. Though the isotropy of the PTFs can be largely improved, these illumination apertures were empirically designed based on empirical criteria related to the shape of the PTF, leaving the underlying theoretical mechanisms unexplored. Furthermore, the frequency responses of the PTFs under these engineered illuminations have not been fully optimized, leading to suboptimal phase contrast and signal-to-noise ratio (SNR) for phase reconstruction.

In this papper, we improve on these works by providing rigorous theory for achieving isotropic DPC, where a new optimal illumination scheme is derived. The major advantages of the new illumination scheme are two-fold. First, it is able to produce a perfectly circularly symmetrical PTF with only 2-axis intensity measurements under partially coherent condition. Thus, it is expected to achieve high-quality phase reconstruction with isotropic transverse resolution and SNR by using only 4 intensity measurements. Second, the resulting PTF achieves a broadband frequency coverage for partially coherent imaging (from 0 to $2NA_{obj}$) with a smooth and significantly enhanced response for both low and high frequencies, which alleviates the ill-posedness of the PTF inversion.

\section{Derivation of optimal illumination scheme}
  Considering a weak scattering object with complex transmission function ${t(r)=e^{1-a(r)+i\phi(r)}}$, it is illuminated by an oblique plane wave with uniform intensity distribution ${S(u_j)}$ (${u_j}$ denotes the spatial frequency of the tilted illumination). Invoking the weak object approximation ${t(r)\approx1-a(r)+i\phi(r)}$ \cite{rose1976nonstandard,hamilton1984improved}, the intensity spectrum of the bright-field image under oblique illumination can be separated into three terms \cite{sun2018single}, including the background, absorption contrast, and phase contrast terms:
\begin{equation}\label{1}
    \begin{split}
       I_{j}(u)=&S({u_j})\delta(u)\left|P(u_j) \right| ^2- \\
       &S({u_j})A(u)\left[{P^*(u_j)P(u+u_j)+P(u_j)P^*(u-u_j)}\right] + \\
       &iS({u_j})\Phi(u)\left[{P^*(u_j)P(u+u_j)-P(u_j)P^*(u-u_j)} \right]  \\
    \end{split}
\end{equation}
where ${P(u)}$ denotes the pupil function of the objective lens (assuming it is an ideal low-pass filter with a cutoff frequency of $\frac{NA_{obj}}{\lambda}$), $\Phi(u)$ is the Fourier transform of the sample's phase function.
In order to generate the phase contrast image $I^{DPC}_{lr}$ (e.g. the left-right DPC) along specific direction of the phase gradient, a pair of images where each $S(u)$ has complementary gradient vector are used to calculated $I^{DPC}_{lr}=\frac{I_r-I_l}{I_r+I_l}$ \cite{mehta2009quantitative,hamilton1984improved}.
Since two illumination patterns are symmetrical along the same axial direction (e.g. x-axis in left-right DPC), the background term and absorption contrast term are cancelled, leaving only the phase contrast term. So the corresponding PTF of left-right DPC can be expressed as:
\begin{equation}\label{2}
        {PTF_{lr}(u)=\frac{\iint{S_{lr}(u_j)\left[{P^*(u_j)P(u+u_j)-P(u_j)P^*(u-u_j)}\right]}d^2u_j}{\iint{|S_{lr}(u_j)||P(u_j)|^2}d^2u_j}}
\end{equation}
It can be found from Eq. (\ref{2}) that once the optical configuration of the microscope is fixed (the pupil function of the objective lens is pre-defined), the PTF is fully determined by the illumination function \cite{sheppard2016interpretation}.

In order to reconstruct the sample's quantitative phase information from Eq. (\ref{2}), we can solve the inverse problem with a single-step deconvolution \cite{tian2015quantitative}. Tikhonov regularization parameter $\beta$  is often introduced in the denominator to avoid singularity in PTF inversion  \cite{bertero1998introduction}:
\begin{equation}\label{3}
        {\phi(r)=\mathcal{F}^{-1}{\left\{\frac{\sum_i\left[{{PTF_i^{*}(u){\cdot}I^{DPC}_i}(u)}\right]}
          {\sum_i|{PTF_i^{*}(u)}|^2+\beta}\right\}}}
\end{equation}
where $PTF_i^{*}(u)$ denotes complex conjugation of PTF along different axial directions. The denominator term  $\sum_i|{PTF_i^{*}(u)}|^2$ represents the synthetic square of amplitude of the multi-axial PTFs, which can be used to indicate the degree of  isotropy for DPC imaging. For simplicity, we use a shorthand notation $C(u)$ for this term in the following analysis.

The following derivation about isotropic DPC is performed in the polar coordinate instead of the Cartesian coordinate; that is, we use $(\rho,\theta)$ instead of $(x,y)$, where $\rho$ and $\theta$ respectively represent the radius and the polar angle. This is because the optical systems have circular symmetry and most circularly symmetric functions are separable in their polar coordinates, i.e., we can write $S_{lr}(\rho,\theta)$ and $S_{ud}(\rho,\theta)$ as a product of two one-dimensional functions about $\rho$ and $\theta$, respectively:
\begin{equation}\label{4}
    \begin{split}
        S_{lr}(\rho,\theta)=L(\rho)M(\theta)\\
        S_{ud}(\rho,\theta)=L(\rho)N(\theta)\\
    \end{split}
\end{equation}
Since phase contrast image is generated by complementary illumination in each axis measurement $S_{lr}=S_{r}-S_{l}$, the illumination function $S_{lr}(\rho,\theta)$ is an odd function about $\theta$. Similarly, the illumination function $S_{ud}(\rho,\theta)$ for the up-down axis is an even function about $\theta$. Thus, $M(\theta)$ and $N(\theta)$ can be further expanded in the Fourier series defined on (-$\pi$, $\pi$]:
\begin{equation}\label{5}
    \begin{split}
        M(\theta)=\sum_{n=1}^\infty{a_nsin(n\theta)}\\
        N(\theta)=\sum_{n=1}^\infty{b_ncos(n\theta)}\\
    \end{split}
\end{equation}

\begin{figure}[htbp]
\centering
\centerline{\includegraphics[width=0.7\linewidth]{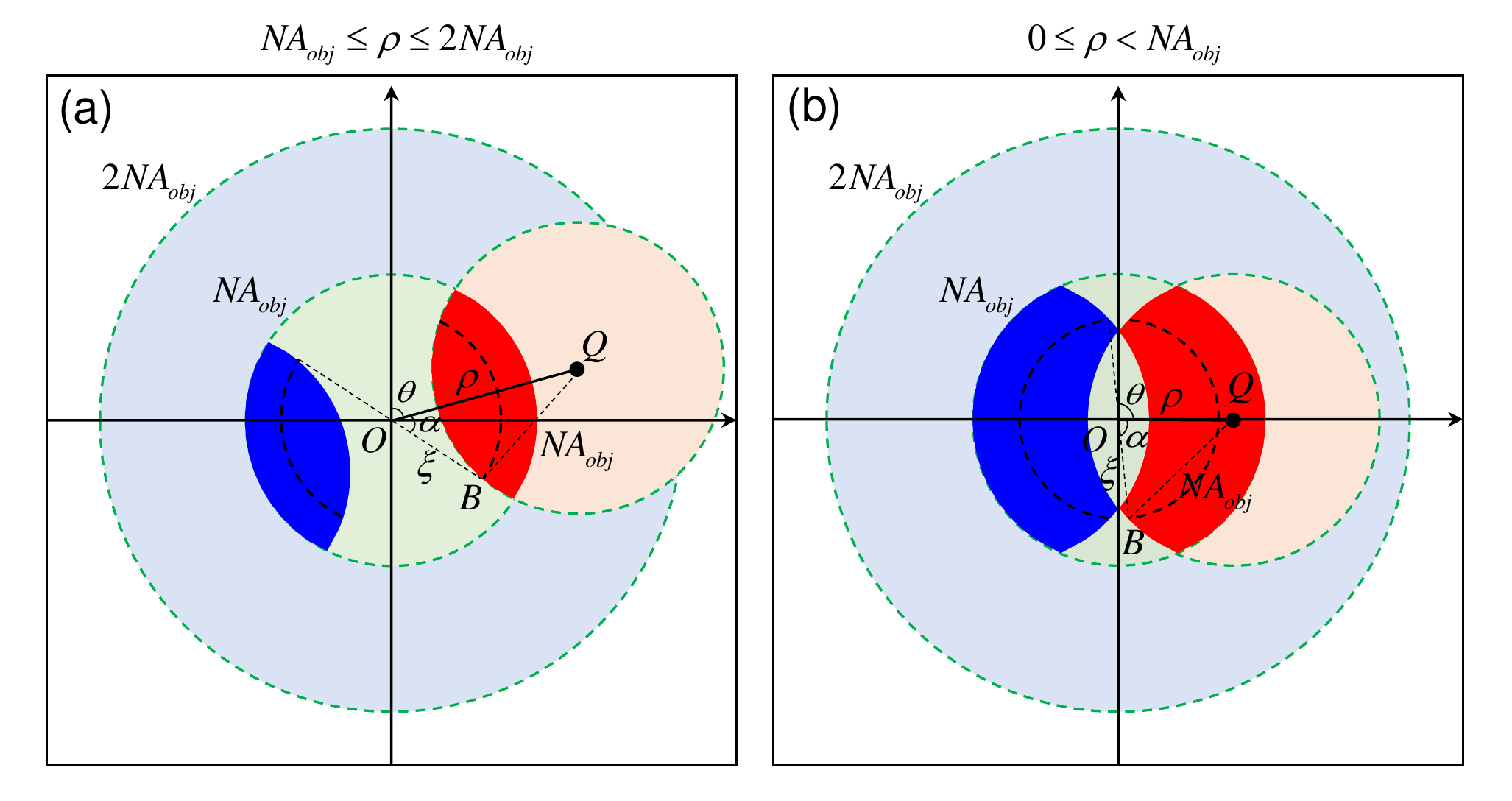}}
\caption{The schematic diagram of the integral for PTF along left-right axis in polar coordinate system. (a) The radius $\rho$ of the point $Q$ is in the range of $NA_{obj}\le\rho\le2NA_{obj}$. (b) The radius $\rho$ of the point $Q$ is in the range of $0 \le\rho< NA_{obj}$.}
\label{Fig1}
\end{figure}

For general illumination and aperture function, the PTF can be calculated by integrating the overlapping areas between the objective pupil function (with its center at the origin $O$) and its off-axis version (with its center at the point $Q$), as illustrated by the red and blue regions in Fig. 1. This is because the illumination falling in the red regions can ensure the point $Q$ is within $P(u+u_j)=1$, and the illumination falling in the blue regions can ensure the point $Q(u)$ is within $P(u-u_j)=1$. However, it should be noted that when the illumination angle is close to the central axis of the objective lens, the two red and blue regions corresponding to $P(u+u_j)$ and  $P(u-u_j)$ will be partially cancelled out by each other. So the integral interval should be partitioned according to the location of point $Q$ [see Figs. \ref{Fig1}(a) and \ref{Fig1}(b)], which can be represented as (taking left-right axis illumination for example)
\begin{equation}\label{6}
PTF_{lr}(\rho,\theta)=\left\{
\begin{array}{rcl}
\frac{2\int_{\rho-NA_{obj}}^{NA_{obj}}\int_{\theta-\alpha}^{\theta+\alpha}{S_{lr}(\xi,\varepsilon)d{\varepsilon}d{\xi}}}{\int_{0}^{NA_{obj}}\int_{0}^{2\pi}{|S_{lr}(\xi,\varepsilon)|}d{\varepsilon}d{\xi}}          & {NA_{obj} \le\rho\le 2NA_{obj}}\\
\frac{2\int_{NA_{obj}-\rho}^{NA_{obj}}\int_{\theta-\alpha}^{\theta+\alpha}{S_{lr}(\xi,\varepsilon)d{\varepsilon}d{\xi}}}{\int_{0}^{NA_{obj}}\int_{0}^{2\pi}{|S_{lr}(\xi,\varepsilon)|}d{\varepsilon}d{\xi}}                     & {0 \le\rho< NA_{obj}}\\
\end{array} \right.
\end{equation}
Substituting Eqs. (\ref{4}) and  (\ref{5}) into Eq. (\ref{6}) gives the PTF along the left-right axis direction. Though the mathematical formulas appear complicated, we should keep in mind that in order to achieve isotropic DPC, it is only required that the synthetic square of amplitude of the multi-axial PTFs $C(\rho,\theta)$ [${C(\rho,\theta)=\left|PTF_{lr}(\rho,\theta)\right|^2+\left|PTF_{ud}(\rho,\theta)\right|^2}$] is only a function of $\rho$. Close inspection of  Eq. (\ref{6}) reveals that the isotropy cannot be achieved if there exists cross trigonometric terms in $C(\rho,\theta)$, suggesting there should be only one single harmonic component in the Fourier series enpension  of Eq. (\ref{5}), that is (see \textbf{Appendix A} for detailed derivation):
 \begin{equation}\label{7}
    \begin{split}
        S_{lr}(\rho,\theta)=L(\rho){sin(n\theta)}\\
        S_{ud}(\rho,\theta)=L(\rho){sin(n\theta)}\\
    \end{split}
      { \qquad(n=1,3,5,...)}
\end{equation}
Note that in Eq. (\ref{7}), we neglect the unimportant constant factors. It should be further noted that when $n$ is even, the illumination pattern is centrosymmetric so that the resulting PTF will be completely canceled out. Therefore, $n$ should be an odd number $(n = 1,3,5,...)$ to guarantee a valid non-zero PTF. Equation (\ref{7}) is the main result of our work, which provides the necessary and sufficient conditions for DPC to achieve perfectly isotropic PTF.

Since no restrictions were imposed on the form of the function $L(\rho)$, $L(\rho)$ can be any function of $\rho$ without affecting the isotropy of the DPC's PTF. Thus, we can optimize the function $L(\rho)$ in order to improve the frequency coverage and response of the corresponding PTF.  In \textbf{Appendix B}, we compare the resultant PTFs of 3 different functions $L(\rho)$: constant weight, linear weight, and Kronecker delta weight functions with a fixed $n = 1$ by simulation. The results shown in Fig. \ref{Fig7} suggest that the delta weight function (a thin annulus) produces the PTF with the highest response at almost all frequencies from 0 to $2NA_{obj}$. In fact,  using an annular source to optimize the PTF of DPC \cite{fan2018wide}, TIE \cite{zuo2017high,li2018optimal}, and FPM \cite{sun2018single,sun2018high} has been demonstrated in our recent studies.

To further analyze the influence of the thickness of the annulus on the PTF, in \textbf{Appendix C}, we further illustrate the corresponding PTFs of different annular thickness by fixing the NA of the outer circles to be $NA_{obj}$ and only changing the thickness of the annulus. As might be expected, the phase contrast is gradually reduced as the annulus width increases, especially at low and high frequencies. This is because the paraxial illumination does not produce low-frequency phase contrast, and only illumination matching the objective $NA_{obj}$ can produce the strong response at all frequencies. When $\sigma\to0$, the phase response finally approaches to that of the constant weight function $L(\rho) = 1$  [Figs. \ref{Fig7}(a1)-\ref{Fig7}(c1) are reproduced]. From the results shown in Figs. \ref{Fig8}(c1) and \ref{Fig8}(d), it can be deduced that we should choose the diameter of the annulus to be equal to that of the objective pupil, and make its thickness as small as possible [$\delta(\rho - NA_{obj})$, where $\delta(\rho)$ is the delta function] to optimize the response of PTF.

Finally, we study the effect of the number $n$ in Eq. (\ref{7}) on the PTF. Three odd $n$ numbers are selected to generate three illumination patterns and corresponding PTFs (See \textbf{Appendix D}). It is shown that, with $n$ increases ($n$ = 3, 5), not only the PTF response is significantly attenuated, but also the number of zero crossings in $C(\rho,\theta)$ increases. In such cases, the reconstruction phase can be severely distorted due to the ill-posedness of the PTF inversion. Thus, $n=1$ provides the optimal PTF for DPC with the strongest response and  no zero crossings from 0 to $2NA_{obj}$. Based on the above analysis, the optimal illumination scheme for isotropic DPC can be represented as:
\begin{equation}\label{8}
    \begin{split}
        S_{lr}(\rho,\theta)=&\delta(\rho - NA_{obj}){sin(\theta)}\\
        S_{ud}(\rho,\theta)=&\delta(\rho - NA_{obj}){cos(\theta)}\\
    \end{split}
\end{equation}
Meanwhile, we can also give the analytical expressions of PTF under the optimal illumination scheme:
\begin{equation}\label{9}
  \begin{split}
        {PTF_{lr}(\rho,\theta)={{sin(\alpha)sin(\theta)}}} \\
      {PTF_{ud}(\rho,\theta)={{sin(\alpha)cos(\theta)}}}
  \end{split}
\end{equation}
Based on the geometric relationship of the isosceles triangle $BOQ$ in Fig. \ref{Fig1}, $\alpha$ is determined by $cos(\alpha)=\frac{\rho}{2NA_{obj}}$. In this case, $C(\rho,\theta)$ can be calculated as:
\begin{equation}\label{10}
        {C(\rho,\theta)=1-\frac{\rho^2}{4{NA_{obj}}^2}}
\end{equation}
Equation (\ref{10}) indicates that $C(\rho,\theta)$ is only related to $\rho$, which means that the corresponding PTF of DPC is isotropic.

\section{Imaging performance of optimal illumination scheme}

\begin{figure}[htbp]
\centering
\centerline{\includegraphics[width=0.8\linewidth]{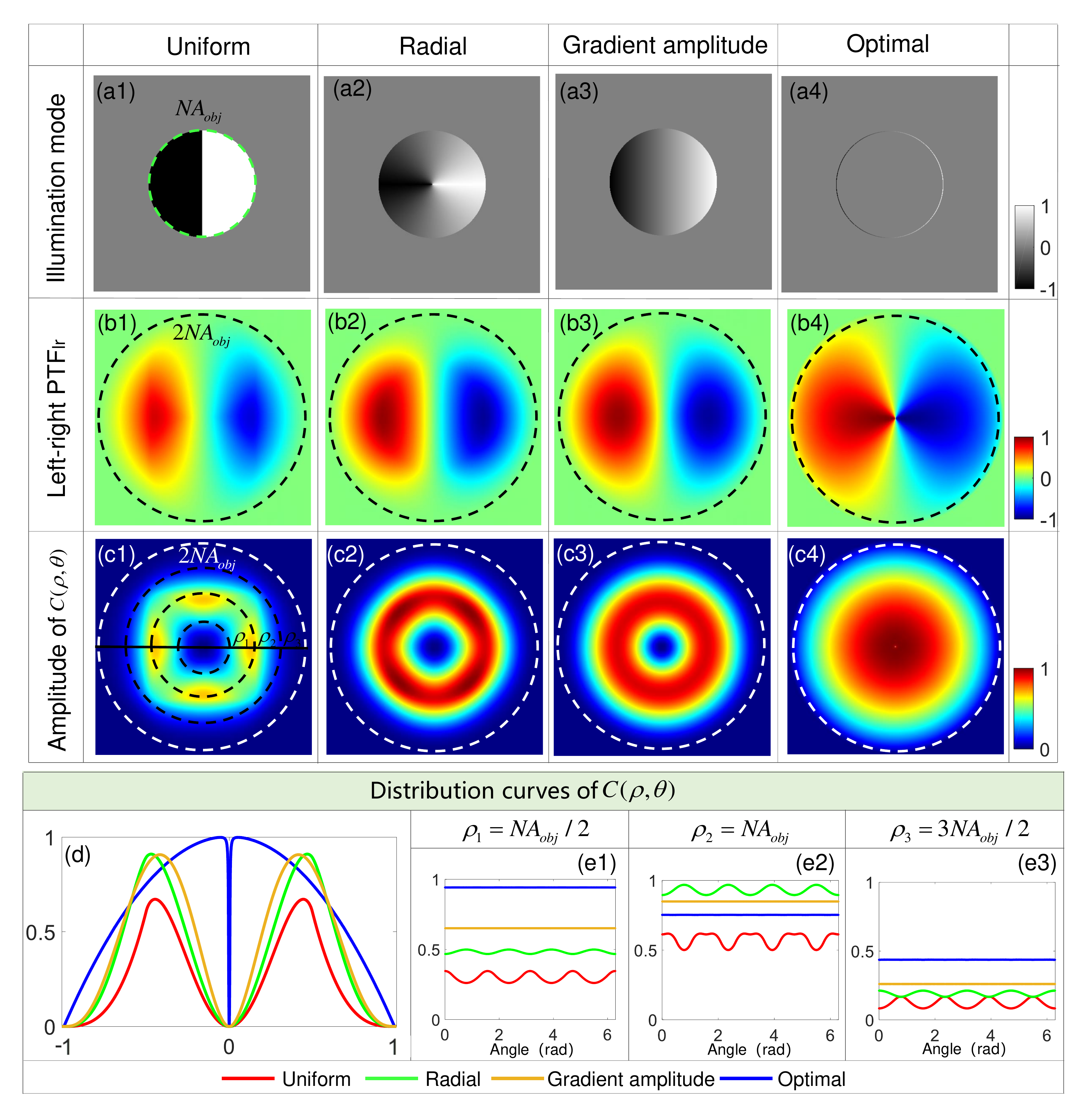}}
\caption{PTF and $C(\rho,\theta)$ with four illumination patterns. (a1)-(a4) Four illumination patterns. (b1)-(b4) PTFs along left-right axis. (c1)-(c4) $C(\rho,\theta)$ with 2-axis illumination. (d) Quantitative curves of $C(\rho,\theta)$ along black straight line under four illumination patterns. (e1)-(e3) Quantitative curves of $C(\rho,\theta)$ under four illumination patterns on three radii.}
\label{Fig2}
\end{figure}

In order to verify the isotropy of optimal illumination scheme, we numerically simulated PTFs and $C(\rho,\theta)$ for four DPC illumination patterns, including 3 state-of-the-arts, namely uniform illumination \cite{tian2015quantitative}, (2-axis) radial illumination \cite{lin2018quantitative}, gradient amplitude illumination \cite{chen2018isotropic}, and the optimal illumination proposed in this work, as shown in Fig. \ref{Fig2}. In Figs. \ref{Fig2}(b1)-\ref{Fig2}(b4), we show PTFs along left-right axis under these different illumination patterns. Obviously, the PTF under optimal illumination scheme has a smooth and significantly enhanced response at almost all frequencies of the theoretical bandwidth of the entire partially coherent imaging (from 0 to $2NA_{obj}$). These can be seen more clearly from the amplitude of $C(\rho,\theta)$ in Figs. \ref{Fig2}(c1)-\ref{Fig2}(c4), where the $C(\rho,\theta)$ under optimal illumination scheme has obviously enhanced values at the low frequency components near the zero frequency and the high frequency components approaching $2NA_{obj}$.

To quantitatively compare these four PTFs, the black cross-section in Fig. \ref{Fig2}(c1) is used to characterize the amplitude of $C(\rho,\theta)$. As shown in Fig. \ref{Fig2}(d), the optimal illumination scheme has the maximum phase contrast at almost all frequencies from 0 to $2NA_{obj}$. Moreover, the peak values of $C(\rho,\theta)$ for the other three illumination patterns are below 0.9, while it can reach 1 under the optimal illumination scheme. The strong phase contrast under the optimal illumination can be finally converted to the quantitative phase images by PTF inversion, resulting in high-quality reconstructions with a uniform background and improved resolution. To compare the degree of isotropy, we further plotted the values along 3 different concentric circles within $C(\rho,\theta)$ in Figs. \ref{Fig2}(e1)-\ref{Fig2}(e3). A constant frequency response can be obtained along the circle under the gradient amplitude illumination and the optimal illumination scheme, suggesting that $C(\rho,\theta)$ obtained under these two illuminations patterns are isotropic. While the uniform illumination \cite{tian2015quantitative} and the radial illumination \cite{lin2018quantitative} cannot generate perfectly isotropic PTF due to their fluctuant frequency responses along the concentric circles.
\begin{figure}[htbp]
\centering
\centerline{\includegraphics[width=0.8\linewidth]{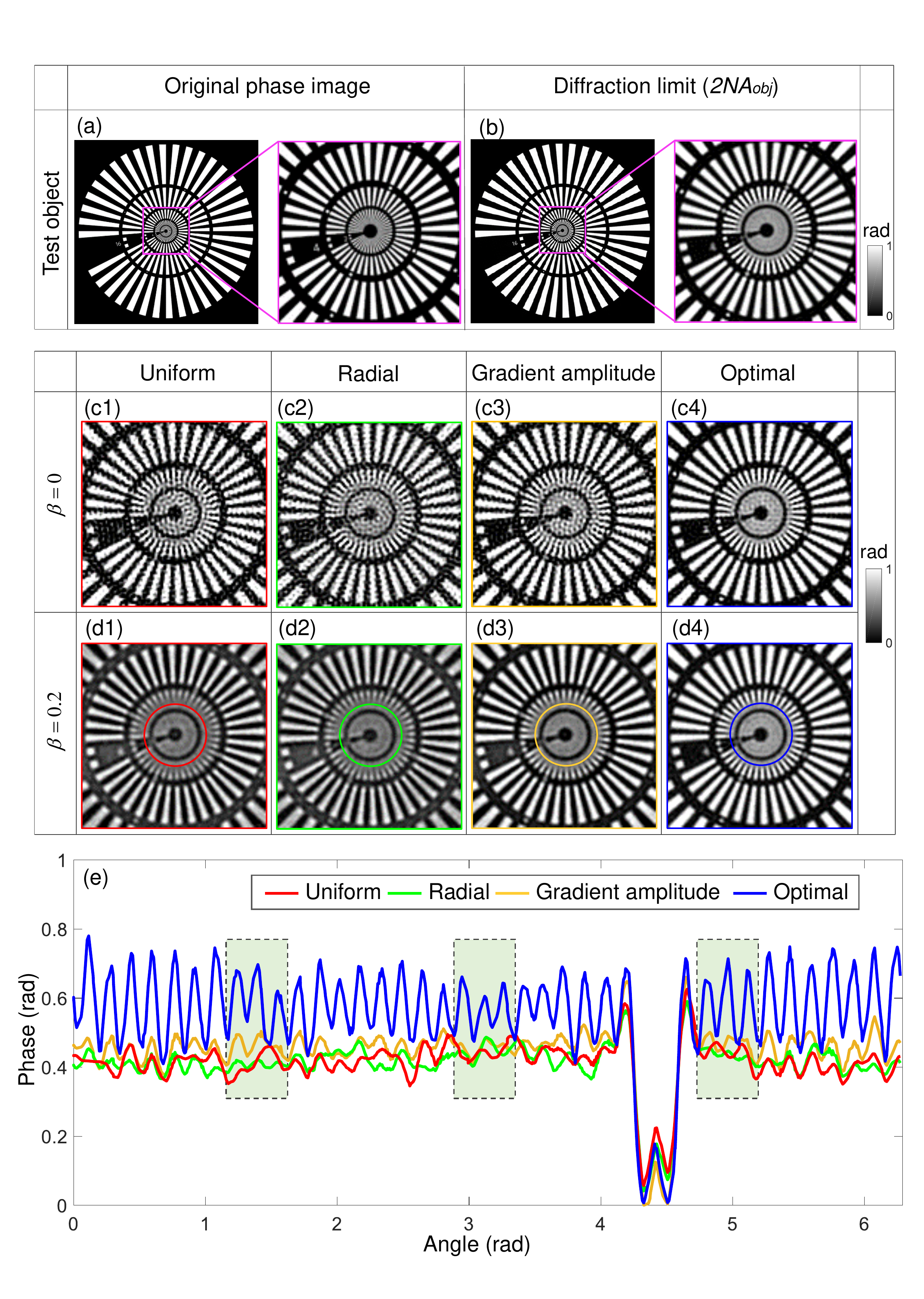}}
\caption{Simulation results with different regularization parameters under four illumination patterns. (a) Original phase image. (b) Diffraction limit phase image of DPC ($2NA_{obj}$). (c1)-(c4), (d1)-(d4) Phase results with regularization parameters of 0 and 0.2. (e) Phase values along a small circle of the same radius in (d1)-(d4) under four illumination patterns.}
\label{Fig3}
\end{figure}

The proposed optimal illumination scheme is then compared with different illumination patterns based on simulations. The Siemens star image [shown in Fig. \ref{Fig3}(a)] \cite{horstmeyer2016standardizing} is used as an example phase object which is defined on a grid with 244 $\times$ 244 pixels with a pixel size of 0.2 $\mu$m $\times$ 0.2 $\mu$m. The wavelength of the illumination is 525 nm, and the $NA_{obj}$ is 0.40. For such an imaging configuration, the best phase imaging resolution can be achieved is 656 nm (${\lambda}/{2NA_{obj}}$), which is also shown in Fig. \ref{Fig3}(b). For partially coherent image calculation, we use the Abbe's method, in which each sub-image corresponding to point source in the aperture plane is superimposed at the image plane to generate captured images of DPC.  To simulate the noise effect, each DPC image is corrupted by Gaussian noise with standard deviation of 0.0003.

In Fig. \ref{Fig3}, we compare the phase retrieval results of different illumination patterns for different regularization conditions. For the case without regularization ($\beta$ is an infinitesimal), the noise corresponding to the frequency components with extremely weak PTF responses was amplified, resulting in grainy artifacts superimposed on the reconstructed phases by using uniform illumination \cite{tian2015quantitative}, (2-axis) radial illumination \cite{lin2018quantitative}, and gradient amplitude illumination \cite{chen2018isotropic} [Figs. \ref{Fig3}(c1)-\ref{Fig3}(c3)]. To stabilize the deconvolution process, a regularization parameter $\beta$ is generally introduced to the denominator of  Eq. (3) to suppress the noise effect. As shown in Figs. \ref{Fig3}(d1)-\ref{Fig3}(d3), when $\beta=0.2$, the grainy artifacts were significantly reduced under these three illumination patterns, but meanwhile, the low-frequency phase values were underestimated and high-frequency features were significantly attenuated. Thus, a more properly chosen regularization parameter is required for these illumination patterns to achieve reliable phase reconstructions under noisy conditions. In contrast, the proposed optimal illumination scheme can always obtain accurate and high-quality reconstructed phases with any regularization parameters attributed to its significantly enhanced PTF response from 0 to $2NA_{obj}$ range [Fig. \ref{Fig2}(b4)], as shown in Figs. \ref{Fig3}(c4) and \ref{Fig3}(d4). To quantitatively compare the resolution of DPC reconstruction results of these four illumination patterns, phase values along a small circle of the same radius in Figs. \ref{Fig3}(d1)-\ref{Fig3}(d4) are extracted and plotted in Fig. \ref{Fig3}(e). As shown in the figure, the green shaded areas corresponding to the frequency components with weak responses in the non-isotropic PTF of the uniform and radial illumination patterns cannot be recovered correctly, while the optimal illumination scheme achieves isotropic resolution as well as much more accurate phase values due to its perfectly isotropic PTF with much stronger responses.

\section{Experimental results}
\noindent
\subsection{Phase resolution target}
\begin{figure}[htbp]
\centering
\centerline{\includegraphics[width=0.9\linewidth]{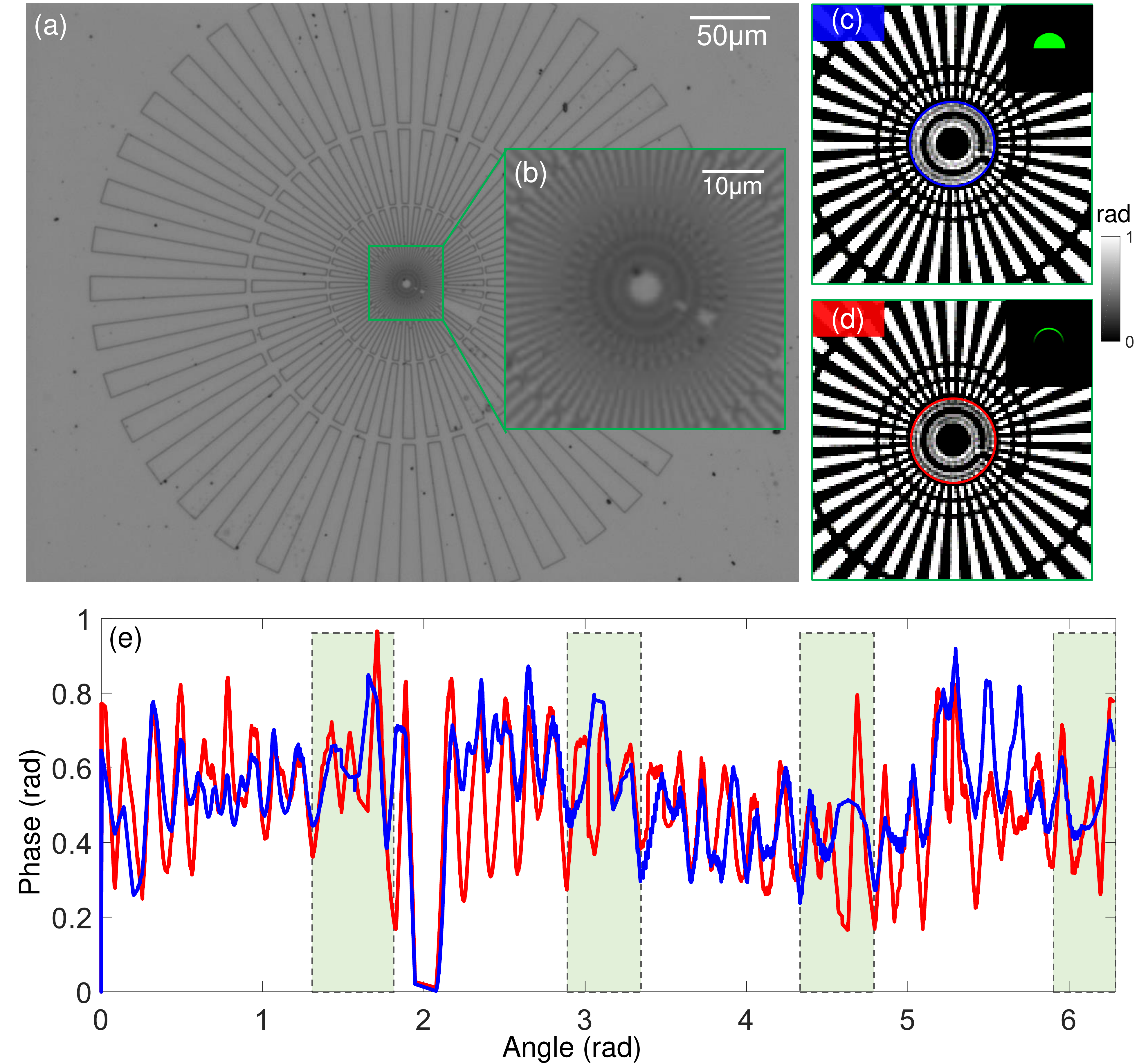}}
\caption{Phase reconstruction results of a phase resolution target $QPT^{TM}$. (a) A bright-field image. (b) Phase reconstruction result under half-circular uniform illumination pattern. (c) Phase reconstruction result under optimal illumination scheme. (d) Phase values along a small circle of the same radius in (c)-(d) under half-circular uniform illumination pattern and optimal illumination pattern.}
\label{Fig4}
\end{figure}

To verify the effectiveness of the optimal illumination scheme experimentally, we firstly measured a pure phase resolution target [Quantitative Phase Microscopy Target ($QPT^{TM}$), Benchmark
Technologies Corporation, USA]. Our setup was built based on a commercial inverted microscope (IX83, Olympus), in which the original condenser diaphragm is replaced by a high contrast anamorphous Silicon (a-Si) thin film transistor (TFT) LCD screen (4.3 inch, pixel resolution 480$\times$272) \cite{zuo2016PCIM}. In our experiments, the build-in halogen white light source with a green interference filter (central wavelength $\lambda$=550 nm, 45 nm bandwidth) was used for illumination, and the LCD screen was used to modulate the illumination in asymmetrical manners. The images were captured by an objective lens with a magnification of $10\times$ and a NA of 0.25 (Olympus PLAN 10X/0.25) and finally digitalized by a CCD camera with pixel size of 3.75 $\mu$m (The imaging source DMK 23U445).
Figure \ref{Fig4}(a) shows a bright-field image captured when the LCD screen is transparent, which has very little intensity contrast. To illustrate the imaging resolution more clearly, a small region near the centre of the image (green-boxed areas) of the reconstructed phases under half-circular uniform illumination and optimal illumination scheme are shown in Figs. \ref{Fig4}(c) and \ref{Fig4}(d), respectively. Furthermore, line profiles of along a small circle of the same radius (blue and red circles) are extracted and illustrated in Fig. \ref{Fig4}(e) to quantitatively compare the highest achievable resolution. These
results are generally consistent with the theoretical prediction as well as our simulation results. It can be seen that the phase details in the green areas, which correspond to the frequency components with weak responses in the non-isotropic PTF of the uniform illumination pattern, cannot be recovered correctly, indicating that half-circular uniform illumination cannot provide phase reconstruction with isotropic transverse resolution and SNR.
Once again, the phase reconstruction result obtained by using the optimal illumination scheme demonstrates much better isotropy.

  \subsection{Unstained hela cells}
\begin{figure}[htbp]
\centering
\centerline{\includegraphics[width=0.9\linewidth]{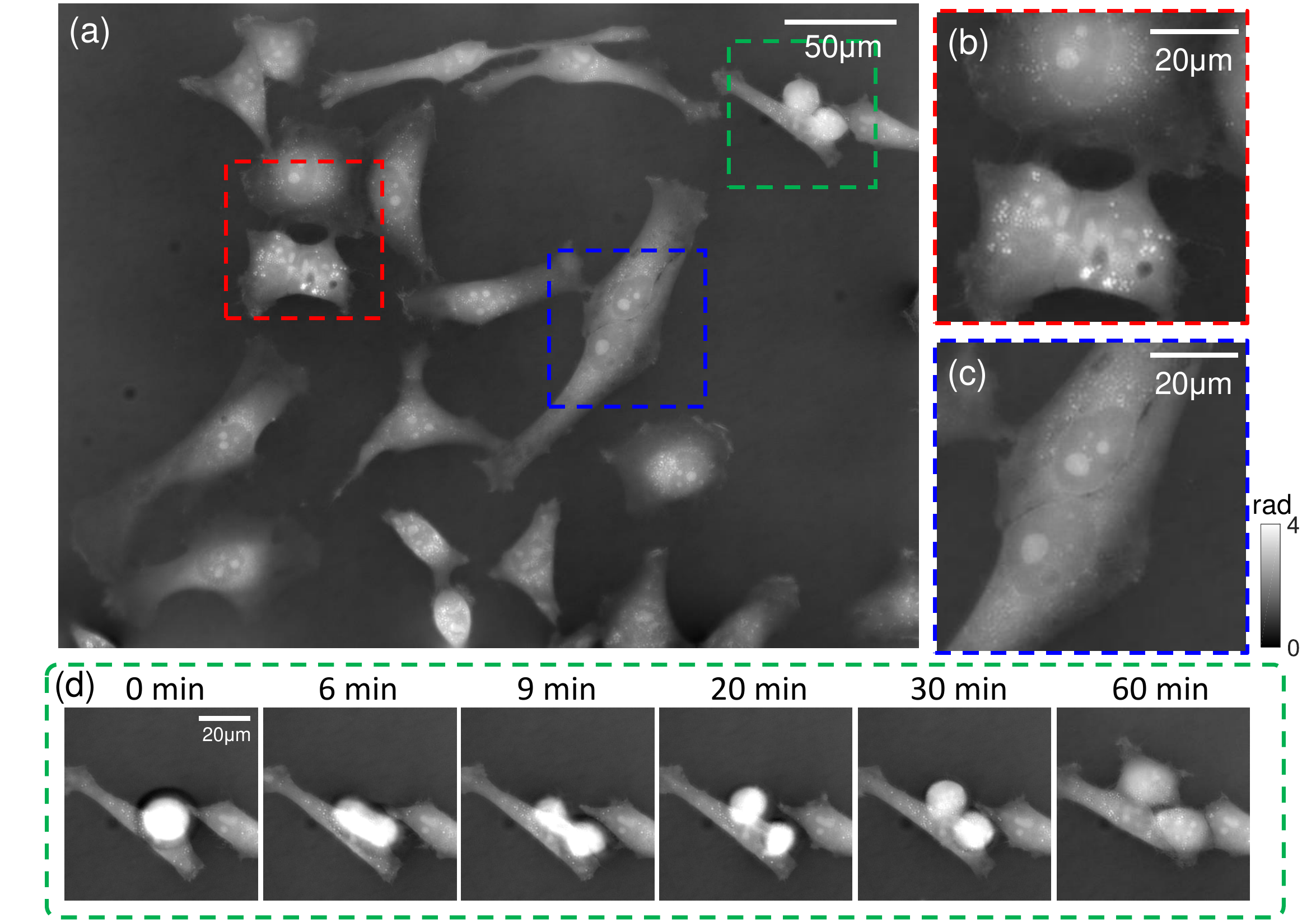}}
\caption{Phase reconstruction results of HeLa cells under optimal illumination scheme. (a) Full field of view phase distribution. (b), (c) Phase maps of two selected zoom. (d) Phase results at different time points.}
\label{Fig5}
\end{figure}

The high-resolution QPI capability of the proposed optimal illumination scheme provides unique possibilities for the label-free imaging of cell growth in culture, using repeated imaging of cultures to assess the progression towards confluence over designated periods of time. In Fig. \ref{Fig5}, we show quantitative phase images of the human cervical adenocarcinoma epithelial (HeLa) cell division process over the course of 5 hours. The experimental setup generally followed the parameters of the previous experiment, except that an objective lens with $10\times, 0.4NA$ (Olympus UPlanSApo 10X/0.4) and an additional $1.25\times$ camera adapter were used (effective magnification $12.5\times$).
A time-lapse movie created with one phase reconstruction per 22 second is provided in \textbf{Visualization 1}.
We show the full field of view phase reconstruction result in this video [one frame from the video is shown in Fig. \ref{Fig5}(a)]. From two selected zooms-in regions (red-boxed and blue-boxed areas) in Figs. \ref{Fig5}(b) and \ref{Fig5}(c), subcellular features, such as cytoplasmic vesicles and pseudopodium, can be clearly observed. In Fig. \ref{Fig5}(d), we further selected one cell [corresponding to the green-boxed region shown in Fig. \ref{Fig5}(a)] to study its morphology during division, which spanned over about 1 hour. These high-resolution phase images clearly reveal the cell morphological changes during different phases of mitosis. Besides, since the optimal illumination scheme requires only two-axis illuminations, all these retracting, extending, reorganizing, migrating, and maturing processes of the cell(s) were recovered accurately without any motion blur. These results demonstrate that optimal illumination scheme is capable of imaging unlabeled cells in a non-invasive manner, allowing for high-resolution QPI over an extended period of time.

 \section{Conclusions and Discussions}
 \noindent
 In this work, the rigorous theoretical analysis about the necessary and sufficient conditions for DPC to achieve perfectly isotropic PTF has been explored. We have derived the optimal illumination scheme to maximize the frequency response and meanwhile achieve isotropic DPC. Compared with traditional DPC method using half-circle illumination, optimal illumination scheme produces a perfectly circularly symmetrical PTF with only 2-axis intensity measurements, avoid missing frequencies and enhances phase response, providing high-quality phase reconstruction with isotropic transverse resolution and SNR. The resultant PFT removes the ill-posedness of the PTF inversion so that artifacts in phase reconstruction results can be significantly reduced. Theoretical analysis, simulations, and experimental results have verified the superiority of our method over state-of-the-arts in both phase reconstruction accuracy and noise-robustness. The investigation of live HeLa cells mitosis \emph{in vitro} has demonstrated that our optimal DPC scheme is a simple, efficient, and stable approach for label-free quantitative cell imaging with subcellular resolution. Furthermore, the intrinsic advantages, such as being non-interferometric, compatibility to bright-field microscopic hardware, and incoherent diffraction-limited resolution up to $2NA_{obj}$, make it a competitive and promising technique for various microscopy applications in life sciences and biophotonics.

 \section*{Appendix A: Derivation of isotropic DPC under generalized illumination conditions}
\begin{figure}[htbp]
\centering
\centerline{\includegraphics[width=0.8\linewidth]{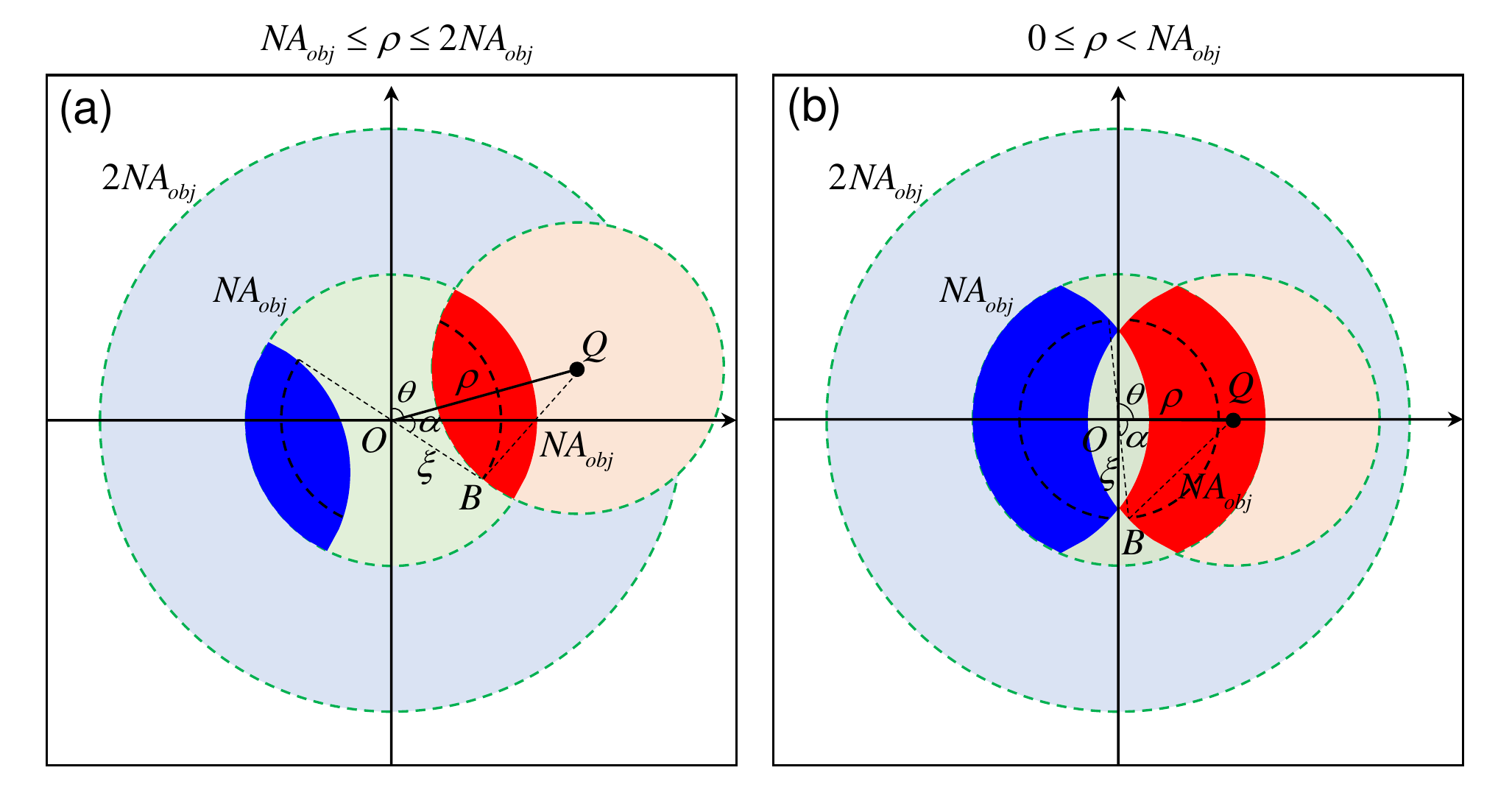}}
\caption{The schematic diagram of the integral for PTF along left-right axis illumination in polar coordinate system.(a) The radius $\rho$ of the point $Q$ is in the range of $NA_{obj}\le\rho\le2NA_{obj}$. (b) The radius $\rho$ of the point $Q$ is in the range of $0 \le\rho< NA_{obj}$.}
\label{Fig6}
\end{figure}

In this Section, we present the rigorous theoretical analysis about the necessary and sufficient conditions for DPC to achieve perfectly isotropic PTF under generalized illumination conditions. As shown in Fig. \ref{Fig6}, we show the schematic diagram of the integral for PTF along left-right axis illumination in polar coordinate system. Since phase contrast image is generated by complementary illumination in each axis measurement $S_{lr}=S_{r}-S_{l}$, the illumination function $S_{lr}(\rho,\theta)$ can be expressed as a periodic odd function about $\theta$. Similarly, we can get illumination function $S_{ud}(\rho,\theta)$ in the up-down axis direction as a periodic even function about $\theta$. Thus,
$S_{lr}(\rho,\theta)$ and $S_{ud}(\rho,\theta)$ can be expanded in the Fourier series defined on $(-\pi, \pi]$:
\begin{equation}\label{11}
    \begin{split}
        S_{lr}(\rho,\theta)=L(\rho)&\sum_{n=1}^\infty{a_{n}{sin(n\theta)}}\\
        S_{ud}(\rho,\theta)=L(\rho)&\sum_{n=1}^\infty{b_{n}{cos(n\theta)}}\\
    \end{split}
     \tag{11}
\end{equation}
where $a_n$ and $b_n$ are a series of constants, $L(\rho)$ is a function about radius $\rho$.

For general illumination and aperture function, the PTF can be calculated by integrating the overlapping areas between the objective pupil function (with its center at the origin $O$) and its off-axis version (with its center at the point $Q$), as illustrated by the red and blue regions in Fig.  \ref{Fig6}. This is because the illumination falling in the red regions can ensure the point $Q$ is within $P(u+u_j)=1$, and the illumination falling in the blue regions can ensure the point $Q(u)$ is within $P(u-u_j)=1$. However, it should be noted that when the illumination angle is close to the central axis of the objective lens, the two red and blue regions corresponding to $P(u+u_j)$ and $P(u-u_j)$ will partially cancel out by each other. So the integral interval should be partitioned according to the location of point $Q$ [see Figs. \ref{Fig6}(a),\ref{Fig6}(b)] which can be represented as (taking left-right axis illumination for example):
\begin{equation}\label{12}
PTF_{lr}(\rho,\theta)=\left\{
\begin{array}{rcl}
\frac{2\int_{\rho-NA_{obj}}^{NA_{obj}}\int_{\theta-\alpha}^{\theta+\alpha}{S_{lr}(\xi,\varepsilon)d{\varepsilon}d{\xi}}}{\int_{0}^{NA_{obj}}\int_{0}^{2\pi}{|S_{lr}(\xi,\varepsilon)|}d{\varepsilon}d{\xi}}          & {NA_{obj} \le\rho\le 2NA_{obj}}\\
\frac{2\int_{NA_{obj}-\rho}^{NA_{obj}}\int_{\theta-\alpha}^{\theta+\alpha}{S_{lr}(\xi,\varepsilon)d{\varepsilon}d{\xi}}}{\int_{0}^{NA_{obj}}\int_{0}^{2\pi}{|S_{lr}(\xi,\varepsilon)|}d{\varepsilon}d{\xi}}                     & {0 \le\rho< NA_{obj}}\\
\end{array} \right.
\tag{12}
\end{equation}
Substituting Eqs. (\ref{11}) into Eq. (\ref{12}) gives the PTF along the left-right axis direction.
\begin{equation}\label{3}
  \begin{split}
PTF_{lr}(\rho,\theta)=&\left\{
\begin{array}{rcl}
\frac{\sum_{n=1}^\infty{a_{n}}sin(n\theta)\int_{\rho-NA_{obj}}^{NA_{obj}}{L(\xi)sin(n\alpha)d{\xi}}}{n\sum_{n=1}^\infty{a_{n}}\int_{0}^{NA_{obj}}{L(\xi)}d{\xi}}          & {NA_{obj} \le\rho\le 2NA_{obj}}\\
\frac{\sum_{n=1}^\infty{a_{n}}sin(n\theta)\int_{NA_{obj}-\rho}^{NA_{obj}}{L(\xi)sin(n\alpha)d{\xi}}}{n\sum_{n=1}^\infty{a_{n}}\int_{0}^{NA_{obj}}{L(\xi)}d{\xi}}      & {0 \le\rho< NA_{obj}}\\
\end{array} \right.\\
PTF_{ud}(\rho,\theta)=&\left\{
\begin{array}{rcl}
\frac{\sum_{n=1}^\infty{b_{n}}cos(n\theta)\int_{\rho-NA_{obj}}^{NA_{obj}}{L(\xi)sin(n\alpha)d{\xi}}}{n\sum_{n=1}^\infty{b_{n}}\int_{0}^{NA_{obj}}{L(\xi)}d{\xi}}          & {NA_{obj} \le\rho\le 2NA_{obj}}\\
\frac{\sum_{n=1}^\infty{b_{n}}cos(n\theta)\int_{NA_{obj}-\rho}^{NA_{obj}}{L(\xi)sin(n\alpha)d{\xi}}}{n\sum_{n=1}^\infty{b_{n}}\int_{0}^{NA_{obj}}{L(\xi)}d{\xi}}      & {0 \le\rho< NA_{obj}}\\
\end{array} \right.\\
\end{split}
\tag{13}
\end{equation}
The above $PTF_{lr}(\rho,\theta)$ and $PTF_{ud}(\rho,\theta)$ are further squared and summed to obtain the synthetic square of amplitude of the multi-axial PTFs by ${C(\rho,\theta)=\left|PTF_{lr}(\rho,\theta)\right|^2+\left|PTF_{ud}(\rho,\theta)\right|^2}$. Isotropic DPC requires $C(\rho,\theta)$ to be a function only about $\rho$, which means that there are no cross trigonometric terms in $C(\rho,\theta)$, suggesting there should be only one single harmonic component in the Fourier series expansion of Eq. (\ref{11}). It should also be noted that when $n$ is even, the illumination pattern is centrosymmetric so that the resulting PTF will be completely canceled out. Therefore, $n$ should be an odd number $(n = 1,3,5,...)$ to guarantee a valid non-zero PTF. As a result, the illumination functions $S_{lr}(\rho,\theta)$ and $S_{ud}(\rho,\theta)$ can be deduced as:
\begin{equation}\label{14}
 \begin{split}
 S_{lr}(\rho,\theta)=&L(\rho)sin(n\theta)\\
 S_{ud}(\rho,\theta)=&L(\rho)cos(n\theta)\\
 \end{split}
 { \qquad(n=1,3,5,...)}
 \tag{14}
\end{equation}
Note that in Eq. (\ref{14}), we neglect the unimportant constant factors. Following this step, the corresponding PTFs can be significantly simplified as follows:
\begin{equation}\label{15}
  \begin{split}
PTF_{lr}(\rho,\theta)=&\left\{
\begin{array}{rcl}
sin(n\theta)\frac{\int_{\rho-NA_{obj}}^{NA_{obj}}{L(\xi)sin(n\alpha)d{\xi}}}{n\int_{0}^{NA_{obj}}{L(\xi)}d{\xi}}          & {NA_{obj} \le\rho\le 2NA_{obj}}\\
sin(n\theta)\frac{\int_{NA_{obj}-\rho}^{NA_{obj}}{L(\xi)sin(n\alpha)d{\xi}}}{n\int_{0}^{NA_{obj}}{L(\xi)}d{\xi}}      & {0 \le\rho< NA_{obj}}\\
\end{array} \right.\\
PTF_{ud}(\rho,\theta)=&\left\{
\begin{array}{rcl}
cos(n\theta)\frac{\int_{\rho-NA_{obj}}^{NA_{obj}}{L(\xi)sin(n\alpha)d{\xi}}}{n\int_{0}^{NA_{obj}}{L(\xi)}d{\xi}}          & {NA_{obj} \le\rho\le 2NA_{obj}}\\
cos(n\theta)\frac{\int_{NA_{obj}-\rho}^{NA_{obj}}{L(\xi)sin(n\alpha)d{\xi}}}{n\int_{0}^{NA_{obj}}{L(\xi)}d{\xi}}      & {0 \le\rho< NA_{obj}}\\
\end{array} \right.\\
\end{split}
\tag{15}
\end{equation}
To further simplify the expression, we define a new function $M(\rho)$ to represent integral term:
\begin{equation}\label{16}
M(\rho)=\left\{
\begin{array}{rcl}
\frac{\int_{\rho-NA_{obj}}^{NA_{obj}}{L(\xi)sin(n\alpha)d{\xi}}}{n\int_{0}^{NA_{obj}}{L(\xi)}d{\xi}}          & {NA_{obj} \le\rho\le 2NA_{obj}}\\
\frac{\int_{NA_{obj}-\rho}^{NA_{obj}}{L(\xi)sin(n\alpha)d{\xi}}}{n\int_{0}^{NA_{obj}}{L(\xi)}d{\xi}}    & {0 \le\rho< NA_{obj}}\\
\end{array} \right.
\tag{16}
\end{equation}
where $\alpha$ is a variable determined by $\rho$.
From the geometric relationship of the triangle BOQ in Fig. \ref{Fig6}, we have $cos(\alpha)=\frac{{\rho}^2+{\xi}^2-{NA_{obj}}^2}{2\rho\xi}$, so $M(\rho)$ here is a function only about $\rho$. Thus $PTF_{lr}(\rho,\theta)$ and $PTF_{ud}(\rho,\theta)$ can be expressed as $PTF_{lr}(\rho,\theta)=sin(n\theta)M(\rho)$ and $PTF_{ud}(\rho,\theta)=cos(n\theta)M(\rho)$. We can get the expression of $C(\rho,\theta)$ as:
\begin{equation}\label{17}
C(\rho,\theta)=M(\rho)^2
\tag{17}
\end{equation}
This result shows that $C(\rho,\theta)$ is independent of $\theta$, suggesting that distribution of $C(\rho,\theta)$ is circularly symmetric. Therefore, we can conclude that the necessary and sufficient conditions for DPC to achieve perfectly isotropic PTF is that the illumination function should be in the form of  $S_{lr}(\rho,\theta)=L(\rho)sin(n\theta)$ and $S_{ud}(\rho,\theta)=L(\rho)cos(n\theta)$, where $L(\rho)$ is an arbitrary function of $\rho$. This conclusion provides great convenience for illumination design of isotropic DPC.
However, it should be noted that Eq. (\ref{16}) generally has no analytical solution, so the PTFs of different patterns can only be evaluated based on numerical simulations.

\section*{Appendix B: Comparison of different illumination schemes for isotropic DPC}
\begin{figure}[htbp]
\centering
\centerline{\includegraphics[width=0.7\linewidth]{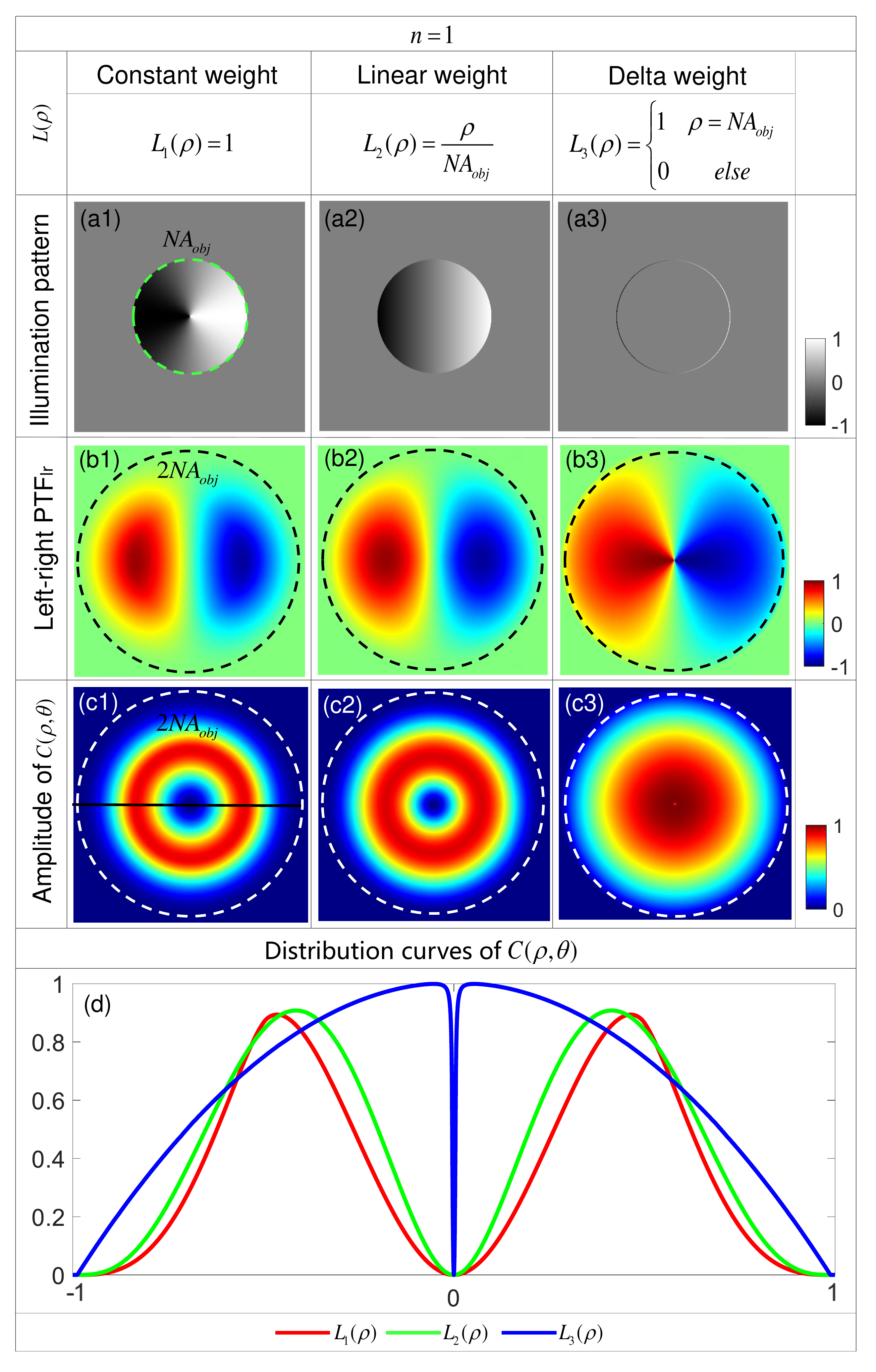}}
\caption{PTF and $C(\rho,\theta)$ with different $L(\rho)$ function. (a1)-(a3) Illumination patterns. (b1)-(b3) PTFs along left-right axis. (c1)-(c3) $C(\rho,\theta)$ with 2-axis illumination. (d) Quantitative curves of $C(\rho,\theta)$ along the black line.}
\label{Fig7}
\end{figure}

Since no restrictions were imposed on the form of the function $L(\rho)$, $L(\rho)$ can be any function of $\rho$ without affecting the isotropy of the DPC's PTF. Thus, we can optimize the function $L(\rho)$ in order to improve the frequency coverage and response of the corresponding PTF. In this section, we compare the PTFs of three different functions $L(\rho)$: constant weight, linear weight, and Kronecker delta weight functions with a fixed $n = 1$ by simulation, as shown in Figs. \ref{Fig7}(a1)-\ref{Fig7}(a3). The PTFs along left-right axis under different illumination patterns are displayed in Figs. \ref{Fig7}(b1)-\ref{Fig7}(b3). It can be seen the PTF with $L_3(\rho)$ produce the PTF with the highest response at almost all frequencies from 0 to $2NA_{obj}$. The results can be seen more clearly from the amplitude of $C(\rho,\theta)$ shown in Figs. \ref{Fig7}(c1)-\ref{Fig7}(c3).
In order to quantitatively characterize the amplitude of these three $C(\rho,\theta)$, the responses along the black cross-section are extracted and compared in Fig. \ref{Fig7}(d). Although these three illumination patterns can all obtain isotropic $C(\rho,\theta)$, PTF corresponding to $L_3(\rho)$ (a thin annulus) has the highest response at almost all frequencies from 0 to $2NA_{obj}$.

\section*{Appendix C: Analysis of influence of annular thickness on the frequency responses of PTF}
\begin{figure}[htbp]
\centering
\centerline{\includegraphics[width=0.7\linewidth]{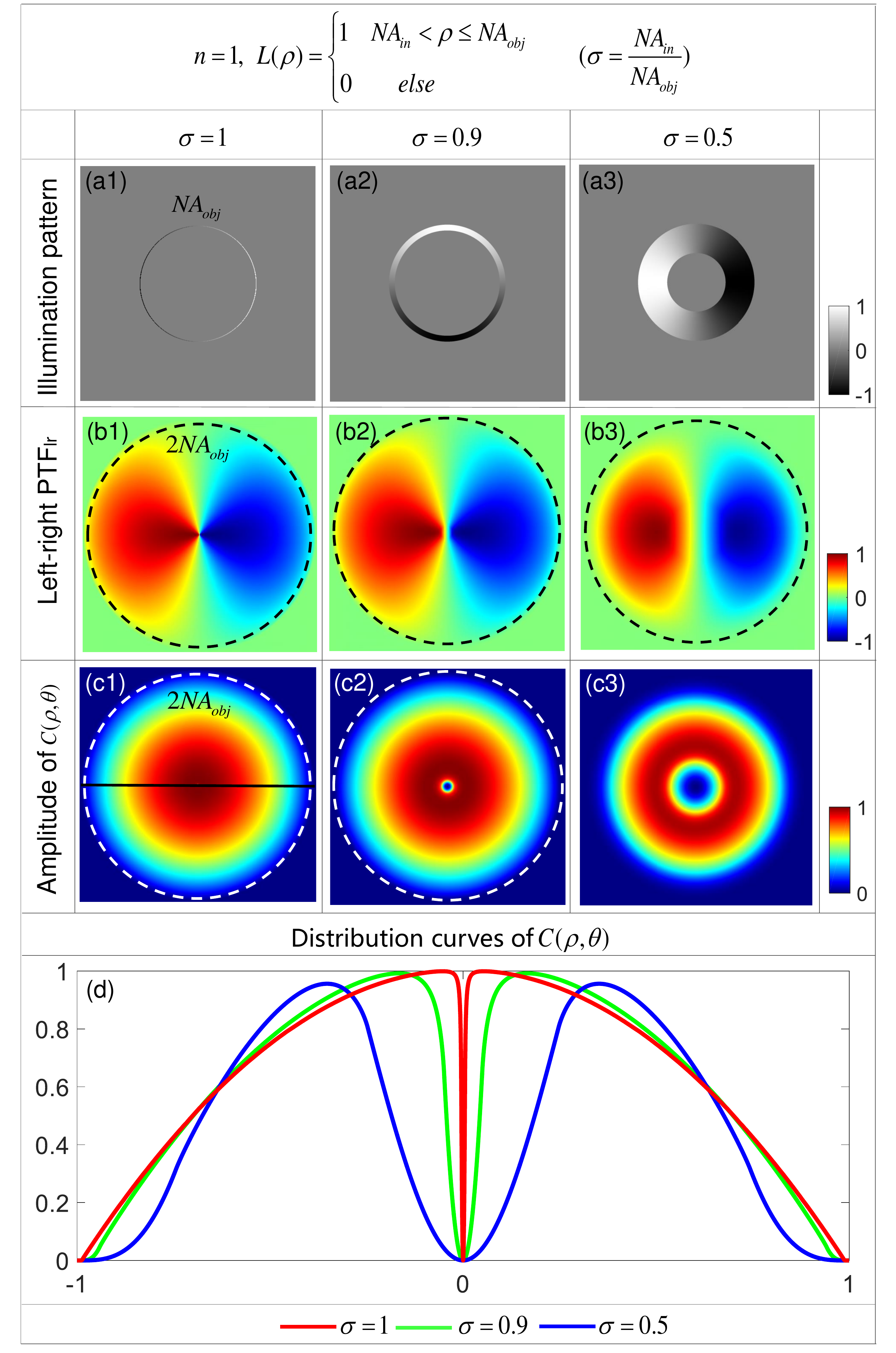}}
\caption{PTF and $C(\rho,\theta)$ with different thickness of the annulus (three $\sigma$). (a1)-(a3) Illumination patterns. (b1)-(b3) PTFs along left-right axis. (c1)-(c3) $C(\rho,\theta)$ with 2-axis illumination. (d) Quantitative curves of $C(\rho,\theta)$ along the black line.}
\label{Fig8}
\end{figure}

We next analyze the effect of the thickness of the annulus on the PTF by fixing the illumination numerical aperture (NA) of the outer circles to be $NA_{obj}$ and only changing the thickness of the annulus ($\sigma=\frac{NA_{ill}}{NA_{obj}}$). As shown in Figs. \ref{Fig8}(b1)-\ref{Fig8}(b3), with the illumination annulus becomes wide, the frequency responses of the PTF are gradually weakened, especially at low and high frequencies. For the synthetic square of amplitude of the multi-axial PTFs $C(\rho,\theta)$ shown in Figs. \ref{Fig8}(c1)-\ref{Fig8}(c3), it can observed that a wider annulus attenuates the response near the zero frequency and the high frequency components approaching $2NA_{obj}$.
When $\sigma\to0$, it can be expected that the phase response finally approaches to that of the constant weight function $L(\rho) = 1$ [Figs. \ref{Fig7}(a1)-\ref{Fig7}(c1) are reproduced]. This is because the paraxial illumination does not produce low-frequency phase contrast, and only illumination matching the objective $NA_{obj}$ can produce the strong response at all frequencies. From the results shown in Fig. \ref{Fig8}, it can be deduced that we should choose the diameter of the annulus to be equal to that of the objective pupil, and make its thickness as small as possible [i.e. the Kronecker delta weight function, $\delta(\rho - NA_{obj})$] to optimize the response of PTF. However, in a practical imaging system, the width of the annulus cannot be infinitesimally thin. From the results shown in Fig. \ref{Fig8}(c1) with Fig. \ref{Fig8}(c2), it can be found that although $C(\rho,\theta)$ corresponding to $\sigma=0.9$ is slightly inferior to that obtained when $\sigma=1$, it can still ensure a relatively strong response over board frequency range while the light throughput can be much improved.

\section*{Appendix D: Analysis of influence of periodicity of the illumination function on the frequency responses of PTF}

\begin{figure}[htbp]
\centering
\centerline{\includegraphics[width=0.7\linewidth]{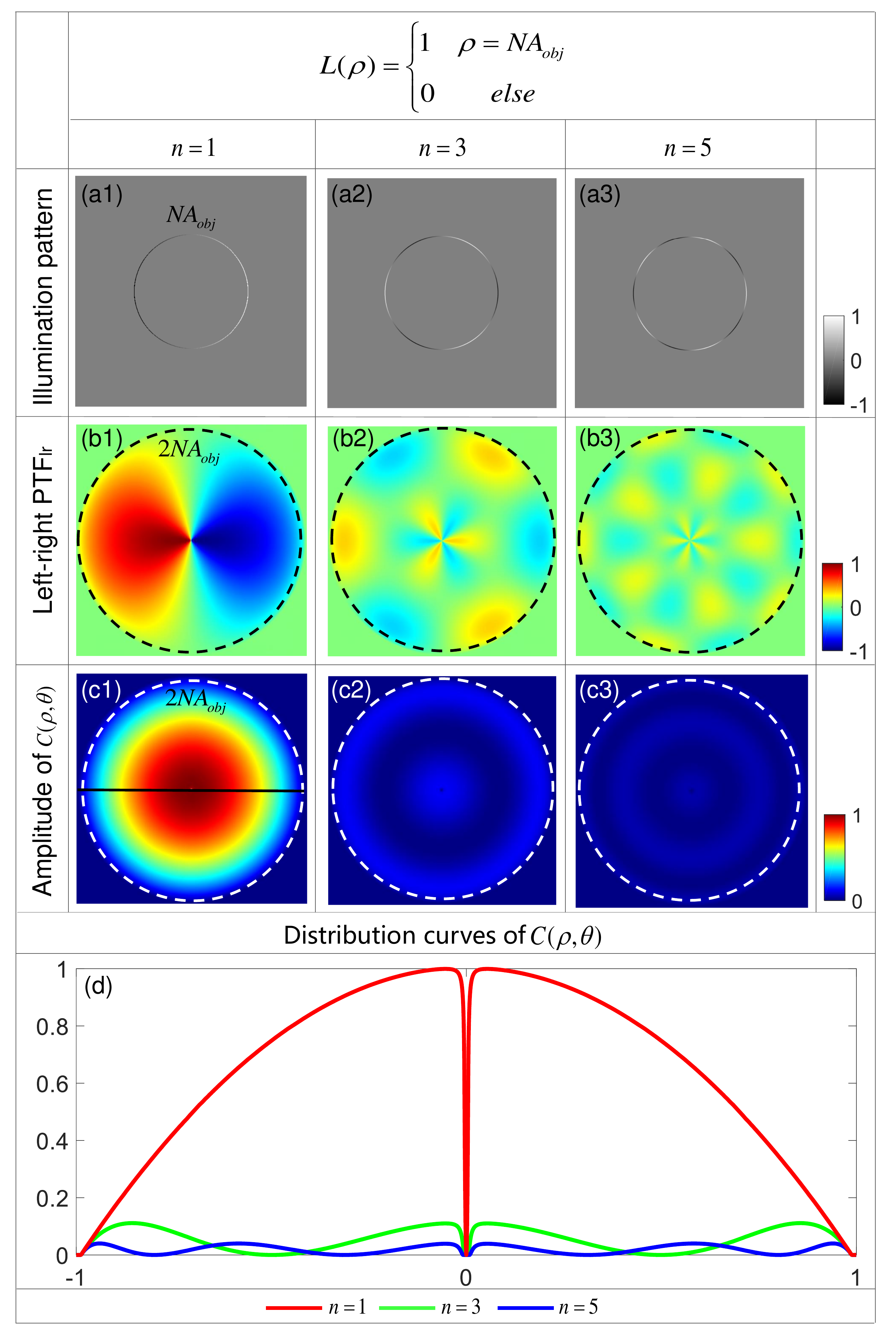}}
\caption{PTF and $C(\rho,\theta)$ with different $n$. (a1)-(a3) Illumination patterns. (b1)-(b3) PTFs along left-right axis. (c1)-(c3) $C(\rho,\theta)$ with 2-axis illumination. (d) Quantitative curves of $C(\rho,\theta)$ along the black line.}
\label{Fig9}
\end{figure}

Finally, in order to the effect of number $n$ in Eq. (\ref{14}) on the PTF, the same $L(\rho)$ function and different $n$ are adopted to generate three illumination patterns, as shown in Figs. \ref{Fig9}(a1)-\ref{Fig9}(a3). The PTF obtained under these three illumination patterns are shown in Figs. \ref{Fig9}(b1)-\ref{Fig9}(b3). It is shown that, with $n$ increases ($n$ = 3, 5), not only the PTF response is significantly attenuated, but also the number of zero crossings in $C(\rho,\theta)$ increases. This is because the increase of $n$ causes more changes of periods for illumination pattern, resulting in a large number of positive and negative apertures to cancel each other out, so that only a PTF with very weak response can be obtained. In such cases, the phase information of the sample is hardly transmitted into the intensity, and the reconstruction phase can be severely distorted due to the ill-posedness of the PTF inversion. Thus, we can conclude that $n=1$ provides the optimal PTF for DPC with the strongest response and  no zero crossings from 0 to $2NA_{obj}$.

\section*{Acknowledgments}
 \noindent
We gratefully acknowledge the support from the National Natural Science Fund of China (61722506, 61505081, 11574152), Final Assembly "13th Five-Year Plan" Advanced Research Project of China (30102070102), Equipment Advanced Research Fund of China (61404150202), National Defense Science and Technology Foundation of China (0106173), Outstanding Youth Foundation of Jiangsu Province of China (BK20170034), The Key Research and Development Program of Jiangsu Province, China (BE2017162), "Six Talent Peaks" project of Jiangsu Province, China (2015-DZXX-009), "333 Engineering" Research Project of Jiangsu Province, China (BRA2016407), Fundamental Research Funds for the Central Universities (30917011204, 30916011322).

%%%%%%%%%%%%%%%%%%%%%%% References %%%%%%%%%%%%%%%%%%%%%%%%%

%%%%%%%%%% If using BibTeX:
\bibliography{sample}

\end{document}